\documentclass[prb,aps,reprint,twocolumn,superscriptaddress]{revtex4-1}

\usepackage{color}
\usepackage{graphicx}
\usepackage{amsmath}
\usepackage{txfonts}
\usepackage{dcolumn}
\begin{document}

\title{Hyperfine coupling constants from internally contracted multireference perturbation theory}
\author{Toru Shiozaki}
\email{shiozaki@northwestern.edu}
\affiliation{\mbox{Department of Chemistry, Northwestern University, 2145 Sheridan Rd., Evanston, IL 60208, USA.}}
\author{Takeshi Yanai}
\affiliation{Department of Theoretical and Computational Molecular Science, Institute for Molecular Science, Okazaki, 444-8585 Aichi, Japan }
\date{\today}

\begin{abstract}
We present an accurate method for calculating hyperfine coupling constants (HFCCs) based on the complete active space second-order perturbation theory (CASPT2) with full internal contraction.
The HFCCs are computed as a first-order property using the relaxed CASPT2 spin-density matrix that takes into account orbital and configurational relaxation due to dynamical electron correlation.
The first-order unrelaxed spin-density matrix is calculated from one- and two-body spin-free counterparts that are readily available in the CASPT2 nuclear gradient program
[M. K. MacLeod and T. Shiozaki, J. Chem. Phys. {\bf 142}, 051103 (2015)], whereas the second-order part is computed directly using the newly extended automatic code generator.
The relaxation contribution is then calculated from the so-called $Z$-vectors that are available in the CASPT2 nuclear gradient program.
Numerical results are presented for the CN and AlO radicals, for which the CASPT2 values are comparable (or, even superior in some cases)
to the ones computed by the coupled-cluster and density matrix renormalization group methods.
The HFCCs for the hexaaqua complexes with V$^\mathrm{II}$, Cr$^\mathrm{III}$, and Mn$^\mathrm{II}$ are also presented to demonstrate the accuracy and efficiency of our code.
\end{abstract}

\maketitle

\section{Introduction}
Analytical nuclear gradients
for fully internally contracted complete-active-space second-order perturbation theory (CASPT2),\cite{Andersson1992JCP,Pulay2011IJQC}
recently developed by one of us,\cite{MacLeod2015JCP,Vlaisavljevich2016JCTC}
can not only allow for geometry optimization of strongly correlated molecules
but also provide the basis for accurate and affordable computations of
properties of such molecules that are sensitive to dynamical electron correlation.
In this work, we develop a method for predicting hyperfine coupling constants (HFCCs) in the electron paramagnetic resonance (EPR) spin Hamiltonian
based on the CASPT2 wave functions.

The hyperfine coupling term in the EPR spin Hamiltonian is
\begin{align}
H^\mathrm{HFC} = \sum_N \sum_{ww'}I_{N,w} A_N^{ww'} S_{w'},
\end{align}
which accounts for coupling between nuclear and electronic spins. The hyperfine interaction causes splitting in EPR spectra
that is often used as an experimental probe for delocalization of unpaired electrons.
Here $w$ and $w'$ label the Cartesian components, $I_{N,w}$ is the nuclear spin vector of nucleus $N$, and $S_{w'}$ is the pseudo-spin operator of the electronic system.
$A_N^{ww'}$ is called the HFCCs formally defined as
\begin{align}
A_N^{ww'} = \frac{\partial^2 E}{\partial I_{N,w} \partial S_{w'}},
\end{align}
where $E$ is the energy of a molecule placed in a magnetic field generated by nuclear spins,
whose vector potential in the Coulomb gauge is given by $g_N \beta_N \mathbf{I}_N \times \mathbf{r}_N/r_N^3$ with
$g_N$ and $\beta_N$ being the nuclear $g$ value and magneton of nucleus $N$, respectively.

Because of its importance in the interpretation of experimental EPR spectra, there have been many developments of {\it ab initio} electronic structure methods for computing HFCCs.\cite{Kauppbook}
For instance, the methods based on single-reference many-body perturbation and coupled-cluster theories
have been studied in the past few decades.\cite{Kristiansen1986JCP,Momose1988JCP,Carmichael1990JPC,Perera1994JCP,Munzarova1999JPCA,Kossmann2010JPCA,Datta2015JCP}
In general, the coupled cluster approaches provide benchmark accuracy when the wave function is well described by a single Slater determinant.
The methods based on the complete active space self-consistent field (CASSCF)
and (uncontracted) multiconfiguration interaction (MRCI) methods have been reported as well.\cite{Fernandez1992JCP,Feller1988JCP,Engels1991CPL,Engels1994JCP}
Yet another approach is based on density functional theory (DFT).\cite{Eriksson1994JCP,Munzarova1999JPCA,Munzarova2000JACS,Neese2003JCP,Arbuznikov2004JCP,Kossmann2007MP}
These approaches have been generalized to relativistic analogues to account for the scalar relativistic and spin--orbit effects.\cite{Komorovsky2006JCP,Autschbach2011JCTC,Malkin2011JCP,Verma2013JCTC,Sharkas2015JCTC}
Recently, Lan {\it et al.}\cite{Lan2014JCTC,Lan2015JCTC} has used the CASSCF method with the density-matrix renormalization group (DMRG) algorithm
to show that
it is capable of reproducing experimental results when used with very large active spaces (up to 36 orbitals).
Though the work by Lan {\it et al.}\cite{Lan2014JCTC,Lan2015JCTC} has clearly shown that the DMRG-CASSCF method provides benchmark accuracy for small systems,
the use of such large active spaces severely limits the scope of applications;
the necessity of such large active spaces originates from the fact that the DMRG-CASSCF model is not designed to efficiently capture dynamical electron correlation,
the importance of which has been emphasized in the previous studies based on the coupled cluster theory.\cite{Carmichael1990JPC,Perera1994JCP,Kossmann2010JPCA,Datta2015JCP}
To overcome this problem, a method based on CASPT2 is developed in this work to efficiently and accurately describe the contributions to HFCCs
from both static and dynamical electron correlation.

The explicit formula for the HFCCs in the absence of spin--orbit coupling is
\begin{align}
A_N^{ww'} = \frac{g_e\beta_e g_N\beta_N}{2S}\sum_{mn} \gamma^{(s)}_{mn}\left(F^{ww'}_{N,mn} - S^{ww'}_{N,mn} \right),
\end{align}
in which $g_e$ and $\beta_e$ are the $g$ value and Bohr magneton of an electron, $g_N$ and $\beta_N$ are those of nucleus $N$, $m$ and $n$ label atomic orbitals, and $\gamma^{(s)}_{mn}$ is a spin-density matrix.
The Fermi-contact and spin--dipole integrals ($F^{ww'}_{N,mn}$ and $S^{ww'}_{N,mn}$, respectively) are defined as
\begin{subequations}
\begin{align}
F^{ww'}_{N,mn} &= \frac{8\pi\delta_{ww'}}{3} \int \phi_m(\mathbf{r}) \delta(\mathbf{r}_N)  \phi_n(\mathbf{r})d\mathbf{r},\\
S^{ww'}_{N,mn} &= \int \phi_m(\mathbf{r}) \frac{3w_Nw'_N - |\mathbf{r}_N|^2 \delta_{ww'}}{|\mathbf{r}_N|^5} \phi_n(\mathbf{r})d\mathbf{r},
\end{align}
\end{subequations}
where $\mathbf{r}_N = \mathbf{r} - \mathbf{R}_N$, $w_N$ is its Cartesian component, and $\phi_m$ is a Gaussian basis function.
In the following, we present an algorithm for computing the unrelaxed and relaxed spin-density matrices from the CASPT2 wave functions, which is followed by numerical examples.

\section{Theory}
The unrelaxed spin-density matrix can be written as
\begin{align}
\gamma^{(s),U}_{xy} = \gamma^{(0s)}_{xy} + \gamma^{(1s)}_{xy} + \gamma^{(2s)}_{xy},
\end{align}
according to the perturbation order, where $x$ and $y$ label any molecular orbitals.
The zeroth-order contribution $\gamma^{(0s)}_{xy}$ is only non-zero in the active--active block and can be readily evaluated from the reference wave function.

The first-order contribution to the spin-density matrix,
\begin{align}
\gamma^{(1s)}_{xy} = (1+\hat{\tau}_{xy})\langle \Phi^{(1)} | a^\dagger_{x\alpha}a_{y\alpha} - a^\dagger_{x\beta}a_{y\beta} | \Phi^{(0)} \rangle,
\end{align}
is calculated from the spin-free density matrices. Here, $\hat{\tau}_{xy}$ is a permutation operator.
For states in their highest spin projection ($m_s = S$),
the spin-density matrices ($\gamma^{(1s)}$) can be expressed in terms of the spin-free density matrices as (in the MO representation)\cite{Luzanov1985TEC,Gidofalvi2009IJQC,Datta2015JCP}
\begin{align}
\gamma^{(1s)}_{xy} = \frac{1}{S+1}\left[\left(\frac{4-N_\mathrm{ele}}{2}\right)\gamma^{(1)}_{xy} - \sum_z \Gamma^{(1)}_{xz, zy}\right],\label{spindensity}
\end{align}
with $N_\mathrm{ele}$ being the number of electrons. The spin-free density matrices on the right hand side
are those appearing in the CASPT2 analytical nuclear gradient theory, i.e.,
\begin{subequations}
\begin{align}
&\gamma^{(1)}_{xy} = (1+\hat{\tau}_{xy})\langle \Phi^{(1)} | \hat{E}_{xy} | \Phi^{(0)} \rangle,\\
&\Gamma^{(1)}_{xy,zw} = (1+\hat{\tau}_{xy}\hat{\tau}_{zw})\langle \Phi^{(1)} | \hat{E}_{xy,zw} | \Phi^{(0)} \rangle,
\end{align}
\end{subequations}
in which $\hat{E}_{xy}$ is a spin-free excitation operator.
In practice, it is convenient to use the following simplified form:
\begin{align}
&(S+1)\gamma^{(1s)}_{xy} = \left\{
\begin{array}{lll}
\displaystyle
-\frac{N_\mathrm{ele,act}}{2}\gamma^{(1)}_{xy} - \sum_r^\mathrm{act} \Gamma^{(1)}_{xr,ry}  && y \in \{ i\}\\
\displaystyle
\left(\frac{4- N_\mathrm{ele,act}}{2}\right)\gamma^{(1)}_{xy} - \sum_r^\mathrm{act} \Gamma^{(1)}_{xr,ry} && y \notin \{ i\}\\
\end{array}
\right.
\end{align}
where $N_\mathrm{ele,act}$ is the number of active electrons, $r$ labels active orbitals, and $\{ i\}$ is the set of closed orbitals.

The same trick, however, cannot be applied to the second-order part, because
the second-order two-body density matrix ($\Gamma^{(2)}$) is not available;
therefore, we directly compute the second-order spin-density matrix,
\begin{align}
\gamma^{(2s)}_{xy} = \langle \Phi^{(1)} | a^\dagger_{x\alpha}a_{y\alpha} - a^\dagger_{x\beta}a_{y\beta} | \Phi^{(1)} \rangle
- S\gamma^{(0s)}_{xy},
\label{spin2}
\end{align}
where $S=\langle \Phi^{(1)} | \Phi^{(1)} \rangle$.
This is done by extending the code generator\cite{MacLeod2015JCP,smith} such that it can be used to compute the ``$\alpha$ density" matrix,
\begin{align}
\gamma^{(2\alpha)}_{xy} = \langle \Phi^{(1)} | a^\dagger_{x\alpha}a_{y\alpha}| \Phi^{(1)} \rangle - S\langle \Phi^{(0)} | a^\dagger_{x\alpha}a_{y\alpha}| \Phi^{(0)} \rangle.
\end{align}
For instance, one of the contributions to the spin-free density matrix,
\begin{align}
& \sum_{aiji'j'rs} T_{ai,rj}T_{ai',sj'} \langle \Phi^{(0)} | \hat{E}_{ai,rj}^\dagger  \hat{E}_{tu} \hat{E}_{ai',sj'} |\Phi^{(0)}\rangle \nonumber\\
& = \sum_{aijrs}T_{ai,rj}(2T_{ai,sj}- T_{aj,si}) \sum_{\rho\sigma}\langle \Phi^{(0)} | a_{r\sigma}  a^\dagger_{t\rho} a_{u\rho} a^\dagger_{s\sigma} | \Phi^{(0)} \rangle \nonumber\\
& = \sum_{aijrs}T_{ai,rj}(2T_{ai,sj}- T_{aj,si}) \nonumber\\
&\quad  \times (2\delta_{su}\delta_{rt} - \delta_{su}\gamma_{tr} - \delta_{tr} \gamma_{su} + 2 \delta_{rs} \gamma_{tu} - \Gamma_{tu,sr}),
\end{align}
is replaced by the following expression,
\begin{align}
&  \sum_{aiji'j'rs}T_{ai,rj}T_{ai',sj'} \langle \Phi^{(0)} | \hat{E}_{ai,rj}^\dagger  a^\dagger_{t\alpha} a_{u\alpha} \hat{E}_{ai',sj'} |\Phi^{(0)}\rangle \nonumber\\
& = \sum_{aijrs}T_{ai,rj}(2T_{ai,sj}- T_{aj,si} ) \sum_{\sigma}\langle \Phi^{(0)} | a_{r\sigma}  a^\dagger_{t\alpha} a_{u\alpha} a^\dagger_{s\sigma} | \Phi^{(0)} \rangle \nonumber\\
& = \sum_{aijrs}T_{ai,rj}(2T_{ai,sj}- T_{aj,si}) \nonumber\\
&\quad  \times (\delta_{su}\delta_{rt} - \delta_{su}\gamma_{\mathbf{tr}} - \delta_{tr} \gamma_{\mathbf{su}} + 2 \delta_{rs} \gamma_{\mathbf{tu}} - \Gamma_{\mathbf{tu},sr}).
\label{spin}
\end{align}
Here, we introduce partially-spin-dependent reference density matrices,
\begin{subequations}
\label{spinden}
\begin{align}
&\gamma_{\mathbf{tu}} = \langle \Phi^{(0)} | a^\dagger_{t\alpha} a_{u\alpha} | \Phi^{(0)} \rangle,\\
&\Gamma_{\mathbf{tu},rs} = \sum_{\rho=\alpha,\,\beta} \langle \Phi^{(0)} | a^\dagger_{r\rho} a^\dagger_{t\alpha} a_{u\alpha} a_{s\rho} | \Phi^{(0)} \rangle,
\end{align}
\end{subequations}
in which only one pair of indices are restricted to the $\alpha$ spin.
The higher-order analogues are likewise defined.
The necessary changes in the code generator are two-fold: First, some of the factors due to spin summation should be halved when the spin-dependent operators participate in the summation
as in the first term of Eq.~\eqref{spin};
second, some of the density matrices are replaced by those in Eq.~\eqref{spinden} and their higher-order analogues.
Note that the active indices in the reduced density matrices have to be sorted in a canonical way such that the spin-dependent indices
can be easily identified.
Finally, we compute the spin-density matrix from the spin-free and $\alpha$ density matrices as $\gamma^{(2s)}_{xy} = 2\gamma^{(2\alpha)}_{xy} - \gamma^{(2)}_{xy}$.
The newly-extended code generator has been validated using the first-order contribution [Eq.~\eqref{spindensity}].
We have also confirmed that the trace of $\gamma_{xy}^{(2s)}$ [Eq.~\eqref{spin2}] is always zero as it should.

The relaxed spin-density matrix is then obtained by adding orbital and configurational relaxation contributions
to the above unrelaxed density matrix using the CASPT2 Lagrangian.\cite{Celani2003JCP,MacLeod2015JCP}
The explicit formula for the relaxed CASPT2 spin-density matrix is
\begin{align}
\gamma^{(s),R}_{xy} = \gamma^{(s),U}_{xy}  + \bar{\gamma}^{(s)}_{xy} + \sum_z \left[\gamma^{(0s)}_{xz} Z_{yz} + Z_{xz}\gamma^{(0s)}_{zy}\right],
\end{align}
in which $Z_{xy}$ is the orbital part of the so-called $Z$-vector.
$\bar{\gamma}_{rs}$ is only nonzero within the active--active block, which reads
\begin{align}
\bar{\gamma}^{(s)}_{rs} = \frac{(1+\hat{\tau}_{rs})}{2} \sum_{I} z_I \langle I|a^\dagger_{r\alpha} a_{s\alpha}- a_{r\beta}^\dagger  a_{s\beta}|\Phi^{(0)}\rangle,
\end{align}
where $z_I$ is the configuration part of the $Z$-vector, and $I$ labels Slater determinants in the active space.
The above $\bar{\gamma}^{(s)}_{rs}$ can be computed by a similar formula to Eq.~\eqref{spindensity}, since the spin-free analogues (both one- and two-body density matrices) are available in the code for solving the $Z$-vector equation.
Regarding the terms in the Lagrangian for the frozen-core approximation,
we note in passing that the frozen-core constraints do not contribute to the relaxed spin-density matrix.
In any event, the frozen-core approximation is typically not used when calculating HFCCs.

\section{Numerical Results}
\begin{table}[tb]
\caption{Isotropic and dipolar HFCCs in MHz for the CN radical ($R=1.1718$~\AA).\label{cntable}}
\begin{ruledtabular}
\begin{tabular}{ldd}
Method & A^\mathrm{iso} & A^\mathrm{dip} \\\hline
\multicolumn{3}{c}{$^{13}$C} \\
CASSCF (9$e$, 8$o$)                     & 629.4 & -51.1 \\
CASSCF (9$e$, 28$o$)\footnotemark[1]\footnotemark[2]  & 596.5 & -52.3 \\
CASSCF (13$e$, 30$o$)\footnotemark[1]\footnotemark[2] & 561.9 & -52.9 \\
CASPT2 (9$e$, 8$o$), unrelaxed          & 692.9 & -53.0 \\
CASPT2 (9$e$, 8$o$), relaxed            & 540.3 & -54.4 \\
CCSD(T)\footnotemark[3]                 & 556.1 & -56.7 \\
B3LYP\footnotemark[1]                   & 572.6 & -59.9 \\
TPSS\footnotemark[1]                    & 504.8 & -59.3 \\
Exp. (Ar matrix)\footnotemark[4]        & 588.1(3) & \multicolumn{1}{c}{$-45(3)$}  \\
\multicolumn{3}{c}{$^{14}$N} \\
CASSCF (9$e$, 8$o$)                     & -19.7 & -17.9 \\
CASSCF (9$e$, 28$o$)\footnotemark[1]\footnotemark[2]  & -3.4 & -19.2 \\
CASSCF (13$e$, 30$o$)\footnotemark[1]\footnotemark[2] & -20.1 & -19.3 \\
CASPT2 (9$e$, 8$o$), unrelaxed          & -15.2 & -17.5 \\
CASPT2 (9$e$, 8$o$), relaxed            & -17.9 & -19.8 \\
CCSD(T)\footnotemark[3]                 & -18.3 & -19.1 \\
B3LYP\footnotemark[1]                   & -18.9 & -21.7 \\
TPSS\footnotemark[1]                    & -16.4 & -22.1 \\
Exp. (Ar matrix)\footnotemark[4]        & -12.6(3) & -15.4(3)
\end{tabular}
\end{ruledtabular}
\footnotetext[1]{Taken from Ref.~\onlinecite{Lan2014JCTC}.}
\footnotetext[2]{The DMRG-CASSCF algorithm was used.}
\footnotetext[3]{Taken from Ref.~\onlinecite{Kossmann2010JPCA}, with a slightly different geometry ($R=1.1555$~\AA).}
\footnotetext[4]{Taken from Ref.~\onlinecite{Easley1969JCP}.}
\end{table}

First, we applied the method to to compute the HFCCs for the CN radical (the bond length was set to $R=1.1718$~{\AA}).
The results are complied in Table~\ref{cntable}.
The EPR-III basis set\cite{Rega1996JCP} was used together with a nearly complete fitting basis set consisting of [$21s21p21d21f21g21h21i$] functions.
The errors due to the density fitting approximation with the above fitting basis set were found negligible to all the digits shown in the table.
The full valence active space was used (9 electrons in 8 orbitals) in the CASPT2 calculation.
The CASPT2 results were compared with the previously reported DMRG-CASSCF results,\cite{Lan2014JCTC}
that computed by CCSD(T),\cite{Kossmann2010JPCA} those from DFT,\cite{Lan2014JCTC} and the experimental values (using Ar matrix).\cite{Easley1969JCP}
Kossmann and Neese used a slightly different bond length ($R=1.1555$~\AA) in their CCSD(T) calculation, though it has been shown that
this difference only affects the HFCCs by up to $0.1$~MHz.\cite{Lan2014JCTC}

The full valence CASPT2 results using the relaxed spin-density matrix agree very well with the CCSD(T) results
with differences up to 2--4~\% for both isotropic and dipolar contributions.
It is apparent from Table~\ref{cntable} that the use of the relaxed spin-density matrix is essential in CASPT2 HFCC calculations; the unrelaxed values are far worse than the CASSCF ones in this case.
The discrepancy between the CASPT2 values and the experimental results should be ascribed to the remaining electron correlation contributions, vibrational contributions,
basis-set incompleteness, and the use of Ar matrix in the experiment.
The wall time for computing the HFCCs using the CASPT2 relaxed density matrix (excluding the CASSCF reference calculations) was
about 4~min using 4 computer nodes, each of which consists of 2 Xeon E5-2650 Sandy Bridge, 2.0~GHz and Infiniband QDR interconnects.
Note that the program is parallelized using MPI3's remote memory access protocol. Threading within a node can still improve.
The results show that the HFCCs computed by CASPT2 are as accurate as those from DMRG-CASSCF using an active space as large as CAS(13$e$, 30$o$)
with a fraction of the computational costs.

\begin{table}[t]
\caption{Isotropic and dipolar HFCCs in MHz for the AlO radical ($R = 1.6176$~\AA).\label{alotable}}
\begin{ruledtabular}
\begin{tabular}{ldd}
Method & A^\mathrm{iso} & A^\mathrm{dip} \\\hline
\multicolumn{3}{c}{$^{27}$Al} \\
CASSCF (9$e$, 8$o$)                     & 830.1 & -47.6 \\
CASSCF (15$e$, 28$o$)\footnotemark[1]\footnotemark[2]   & 629.3 & -53.1 \\
CASSCF (15$e$, 33$o$)\footnotemark[1]\footnotemark[2]   & 573.1 & -54.8 \\
CASSCF (21$e$, 36$o$)\footnotemark[1]\footnotemark[2]   & 712.7 & -54.2 \\
CASPT2 (9$e$, 8$o$), unrelaxed          & 998.8 & -53.3 \\
CASPT2 (9$e$, 8$o$), relaxed            & 788.3 & -55.8 \\
B3LYP\footnotemark[1]                   & 512.2 & -60.0 \\
TPSS\footnotemark[1]                    & 656.8 & -56.1 \\
CCSD(T)\footnotemark[3]                 & 565.3 & -56.2 \\
Exp. (gas phase)\footnotemark[4]        & 738.0(14) & -56.4(1) \\
Exp. (Ne matrix)\footnotemark[5]        & \multicolumn{1}{c}{766(2)} & -53.0(7) \\
\multicolumn{3}{c}{$^{17}$O} \\
CASSCF (9$e$, 8$o$)                     & -1.3 & 37.3 \\
CASSCF (15$e$, 28$o$)\footnotemark[1]\footnotemark[2]   & -42.3 & 49.2 \\
CASSCF (15$e$, 33$o$)\footnotemark[1]\footnotemark[2]   & -57.3 & 55.5 \\
CASSCF (21$e$, 36$o$)\footnotemark[1]\footnotemark[2]   & -35.0 & 52.2 \\
CASPT2 (9$e$, 8$o$), unrelaxed          & -0.5 & 34.4 \\
CASPT2 (9$e$, 8$o$), relaxed            & 13.4 & 52.6 \\
B3LYP\footnotemark[1]                   & 8.2 & 66.2 \\
TPSS\footnotemark[1]                    & 9.5 &  59.9 \\
CCSD(T)\footnotemark[3]                 & 19.3 & 58.9 \\
Exp. (Ne matrix)\footnotemark[5]        & 2 & 50
\end{tabular}
\end{ruledtabular}
\footnotetext[1]{Taken from Ref.~\onlinecite{Lan2014JCTC}.}
\footnotetext[2]{The DMRG-CASSCF algorithm was used.}
\footnotetext[3]{Taken from Ref.~\onlinecite{Kossmann2010JPCA}.}
\footnotetext[4]{Taken from Ref.~\onlinecite{Yamada1990JCP}.}
\footnotetext[5]{Taken from Refs.~\onlinecite{Knight1971JCP} and \onlinecite{Knight1997JCP}.}
\end{table}

\begin{table}
\caption{Isotropic transition-metal HFCCs for $^4$[V(H$_2$O)$_6$]$^{2+}$, $^4$[Cr(H$_2$O)$_6$]$^{3+}$, and $^6$[Mn(H$_2$O)$_6$]$^{2+}$ in MHz.\label{metaltable}}
\begin{ruledtabular}
\begin{tabular}{lddd}
Method & \multicolumn{1}{c}{$^4$[V(H$_2$O)$_6$]$^{2+}$} & \multicolumn{1}{c}{$^4$[Cr(H$_2$O)$_6$]$^{3+}$} & \multicolumn{1}{c}{$^6$[Mn(H$_2$O)$_6$]$^{2+}$} \\
\hline
UB3LYP\footnotemark[1] & -162 & 32 & -163 \\
UBP\footnotemark[1] & -166 & 33 & -164 \\
ROHF   & 0.0 & 0.0 & 0.0 \\
CASSCF (10$o$) & -0.6 & -23.1 & 66.8 \\
CASSCF (14$o$) & -126.1 & 4.2 & -106.3 \\
CASSCF (19$o$)\footnotemark[2] & -120.5 & 4.1 & -117.2\\
CASSCF (23$o$)\footnotemark[2] & -132.7 & 8.9 & -113.0\\
CASPT2 (ROHF)  & -268.9 & 56.8 & -301.6 \\
CASPT2 (10$o$) & -232.0 & 45.3 & -224.2 \\
$\Delta$SO\footnotemark[3] & -10 & 3 & -2 \\
Exp.\footnotemark[4]  & -247 & 55 & -245
\end{tabular}
\end{ruledtabular}
\footnotetext[1]{Data excluding spin--orbit contributions, taken from Ref.~\onlinecite{Neese2003JCP}.}
\footnotetext[2]{Calculated using the DMRG-CASSCF algorithm.}
\footnotetext[3]{Spin--orbit contributions calculated in Ref.~\onlinecite{Neese2003JCP} using UB3LYP.}
\footnotetext[4]{See main text for the references.}
\end{table}

Next, we computed the HFCCs for the AlO radical ($R=1.6176$~\AA).
The IGLO-III and EPR-III basis sets were used for Al and O, respectively,\cite{Rega1996JCP,KutzelniggIGLObook}
and the active space was chosen to be the full valence space consisting of 9 electrons in 8 orbitals.
The same nearly-complete fitting basis set was used, as described above.
The CASPT2 results were compared to those obtained by the DMRG-CASSCF,\cite{Lan2014JCTC}
CCSD(T),\cite{Kossmann2010JPCA} and DFT\cite{Lan2014JCTC} and the experimental values.\cite{Knight1971JCP,Yamada1990JCP,Knight1997JCP}
The previous computations were performed using the same basis set.
The results have been tabulated in Table~\ref{alotable}.
Since the ground state of the AlO radical has a multi-configuration character due to the resonance between Al$^+$O$^-$ and Al$^{2+}$O$^{2-}$,
single-reference electron correlation methods, such as CCSD(T), fail to describe the spin-density matrix accurately, resulting in errors in the predicted HFCCs up to 170~MHz.\cite{Kossmann2010JPCA}
The values obtained by CASPT2 for the Al and O centers are comparable and superior to the largest DMRG-CASSCF computation reported in Ref.~\onlinecite{Lan2014JCTC}, respectively,
albeit much reduced computational costs.
The deviations from the experimental values on the Al center are only 7~\% and 1~\% for the isotropic and dipolar contributions, respectively.
The deviations  on the O center are 11 and 3~MHz.
It is again observed that the use of the relaxed spin-density matrix is crucial for accurate computation of HFCCs.

\begin{figure}
\includegraphics[keepaspectratio,width=0.48\textwidth]{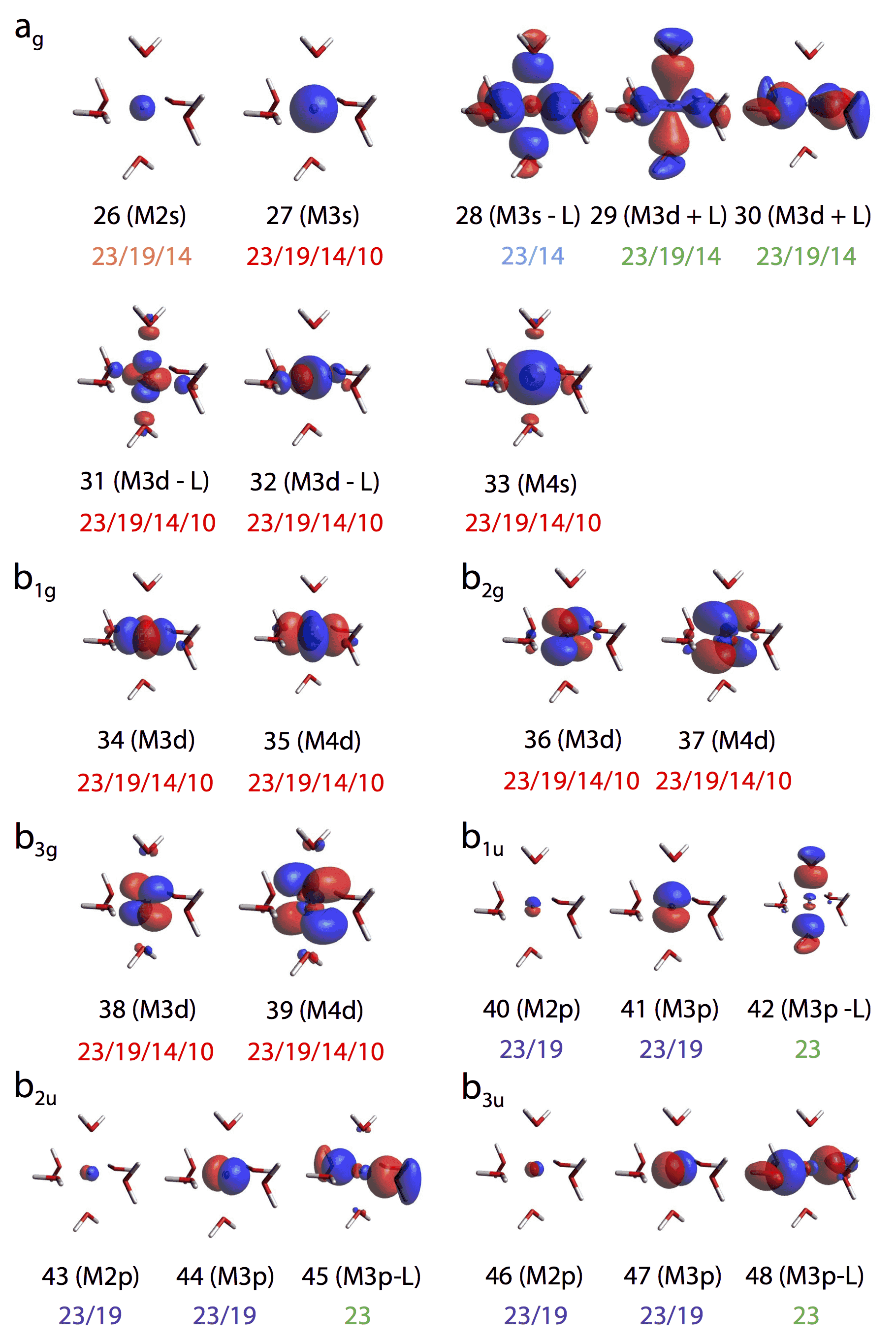}
\caption{Active orbitals used in the calculations of the transition-metal HFCCs for M(H$_2$O)$_6$ (M $=$ V$^\mathrm{II}$, Cr$^\mathrm{III}$, and Mn$^\mathrm{II}$).\label{active}}
\end{figure}

Finally, we applied our method to the isotropic transition-metal HFCCs for $^4$[V(H$_2$O)$_6$]$^{2+}$, $^4$[Cr(H$_2$O)$_6$]$^{3+}$, and $^6$[Mn(H$_2$O)$_6$]$^{2+}$.
Note that the anisotropic HFCCs are zero due to symmetry.
The molecular structures were determined at the UB3LYP-D3BJ/6-31G** level of theory (the geometric data are given in the supporting information).
The active orbitals are shown in Fig.~\ref{active}; 10 orbitals were included in the CASPT2 calculations, consisting of
3s, 4s, and 3d orbitals and t$_\mathrm{2g}$ 4d orbitals of the transition metals, some of which were mixed with ligand orbitals.
The CASSCF calculations were performed using the {\sc orz} package of Yanai and co-workers,
and the resulting orbitals were read via the {\sc molden} interface in {\sc bagel} and used in the CASPT2 calculation.
The triply polarized [CP(PPP)] basis set\cite{Neese2002ICA} and the def2-SVP basis set\cite{Eichkorn1997TCA} were used for the transition metals and ligands, respectively.
The def2-TZVPP-JKFIT basis set\cite{Weigend2008JCompC} was used for density fitting in the CASPT2 calculations.
The errors due to density fitting were found to be negligible.

The calculated isotropic HFCCs are shown in Table~\ref{metaltable}, together with the experimental values taken from Ref.~\onlinecite{Neese2003JCP}
(the original experimental works were reported in Refs.~\onlinecite{McGarveyChapter} and \onlinecite{Upreti1974JMR}, from which these values were derived).
The results from DFT\cite{Neese2003JCP} and large-scale CASSCF based on the DMRG algorithm with $M=512$ are also shown.
The CASPT2 values based on the ROHF (3 active orbitals for the V$^\mathrm{II}$ and Cr$^\mathrm{III}$ complexes and 5 active orbitals for the Mn$^\mathrm{II}$ complex)
and 10-orbital CASSCF references are both in good agreement with the experimental values; the former slightly overestimates the magnitude of the HFCCs
(up to 50~MHz for $^6$[Mn(H$_2$O)$_6$]$^{2+}$), while
 the latter is consistently within 20~MHz from the experimental values.
The spin--orbit contributions estimated by DFT in Ref.~\onlinecite{Neese2003JCP} suggest that the agreement would improve if the results are corrected for the spin--orbit effect.
It is worth noting that the CASPT2 method provides excellent accuracy without breaking the spin symmetry.
For these complexes, spin-unrestricted B3LYP severely underestimate the HFCCs values.
The DMRG-CASSCF results, even with 23 orbitals in the active space, underestimate the magnitudes of the metal HFCCs by 50--130 MHz,
which clearly shows that the explicit treatment of dynamical correlation is necessary for accurate simulation of the HFCCs.
The wall times for computing the CASPT2 values are around 2 min on 16 CPU cores (2 Xeon E5-2650 CPUs) with the ROHF reference and
15 min on 64 CPU cores (8 Xeon E5-2650 CPUs) with the 10-orbital CASSCF reference, respectively.

\section{Conclusions}
In this work, we have developed an algorithm for computing the HFCCs from the CASPT2 spin-density matrices including orbital and configurational relaxation and implemented it into
and an efficient computer program.
The program is an extension of the analytical nuclear gradient code for fully internally contracted CASPT2.\cite{MacLeod2015JCP,Vlaisavljevich2016JCTC}
In our implementation, spin-density matrices can be computed as efficiently as spin-free analogues, owing to the use of partially spin-dependent density matrices.
We have applied this approach to the CN and AlO radicals and the hexaaqua complexes with V$^\mathrm{II}$, Cr$^\mathrm{III}$, and Mn$^\mathrm{II}$
to demonstrate the accuracy and efficiency of our program.
The code has been implemented in the {\sc bagel} package,\cite{bagel} which is publicly available under the GNU General Public License.
The implementation in the {\sc orz} package is under way.
To account for relativistic contributions to HFCCs,
generalization of the present work to the fully relativistic framework based on the four-component Dirac equation
will be investigated in the future.

\begin{acknowledgments}
T.S. has been supported by the National Science Foundation CAREER Award (CHE-1351598).
The development of the underlying program for the CASPT2 nuclear gradients was in part supported
by the Air Force Office of Scientific Research Young Investigator Program (AFOSR Grant No.~FA9550-15-1-0031).
T.Y. acknowledges support by JSPS KAKENHI Grant Numbers JP16H04101 and JP15H01097.
\end{acknowledgments}

\end{document}